\documentclass{amsart}        

\usepackage{amssymb,latexsym}
\usepackage{graphicx}
\usepackage{subfig}
\usepackage{algorithmic}

\begin{document}
\title{Decting Errors in Reversible Circuits With Invariant Relationships}
\author{
  Nuno Alves \\
  nuno@brown.edu \\
  Engineering Division, Brown University
}

\begin{abstract}
Reversible logic is experience renewed interest as we are approach the limits of CMOS technologies. While physical implementations of reversible gates have yet to materialize, it is safe to assume that they will rely on faulty individual components. In this work we present a present a method to provide fault tolerance to a reversible circuit based on invariant relationships. 
\end{abstract}

\maketitle

\vspace*{-2mm}
\section{Introduction}\label{sec:introduction}

The advent of CMOS scalability and power dissipation limits is bringing a renewed interest in reversible architectures. In essence, these architectures contain logic gates that are able to reuse energy that was spent to generate previous results. In fact, the main purpose of reversible architectures is to improve the computational efficiency further than the fundamental Von Neumann limit ~\cite{Neumann66} of \emph{kTln2} energy dissipated per irreversible bit operation, where \emph{k} is Boltzmann's constant and \emph{T} is absolute temperature. This limit is characterized as fundamental because its value is independent of the properties of the device, materials, or circuit that may be used to implement the logic operation ~\cite{meindl2000}. In theory, a fully reversible architecture would dissipate no energy.

\subsection{Historical Perspective}
In 1961, Rolf Landauer showed that traditional binary (irreversible) gates lead to power dissipation regardless of their implementation~\cite{Landauer61}. He did so by noticing that any logically irreversible manipulation of information, such as the erasure of a bit or the merging of two computation paths, always increase the  entropy of the system. For each computational operation in which a single bit of logical information is lost, the entropy generated is at least \emph{kTln2}, which means the system will have an energy increase of \emph{kTln2} joules. In 1973, Charles Bennet demonstrated that unlike the erase operation, the read operation did not bring any entropy increase to the system~\cite{Bennett73}. Therefore, one possible way to circunvent the release of energy into the system per operation, is to have a circuit built from gates that are capable of using previous logic outputs as inputs to that same gate. In other words, these gates need to be reversible.

\subsection{Synthesis of Reversible Gates}
In accordance to Bennett's demonstration, a reversible gate is one that: 
\begin{enumerate}
\item{Has an equal number of inputs and outputs}
\item{Each input/output vector must be unique}
\end{enumerate}

\begin{figure}
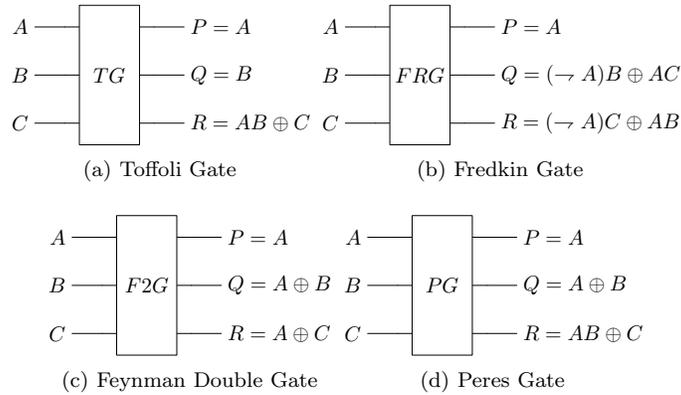

  \centering
  \subfloat[Toffoli Gate]{\includegraphics[scale=0.8]{paper.1}}        
  \subfloat[Fredkin Gate]{\includegraphics[scale=0.8]{paper.2}}    
  
  \subfloat[Feynman Double Gate]{\includegraphics[scale=0.8]{paper.3}}
  \subfloat[Peres Gate]{\includegraphics[scale=0.8]{paper.4}}         
  
  \caption{Toffoli~\cite{tt1980}, Fredkin~\cite{Fredkin82conservativelogic}, Peres~\cite{Peres85} and Feynman~\cite{Feynman85quantummechanical} reversible gates}
  \label{fig:sampleRevGates}
\end{figure}

When different input vectors have the same output vector, the function is irreversible. It is irreversible because it is impossible to determine the input vectors from the output vectors. The NOT gate is reversible, as every single output maps to a unique input, whereas the NAND and XOR gates are not. Several reversible gates have been proposed. Some well known gates are listed on figure~\ref{fig:sampleRevGates}. Of the listed gates all are universal, since combinations of them can be used to accomplish any arbitrary Boolean function. For example, in the Toffoli gate on figure~\ref{fig:sampleRevGates} when C=1, R=$\rightharpoondown$(AB), which is the universal Boolean operation NAND.

\begin{figure}
  \centering
\begin{tabular} {|ccc|ccc|}  \cline{1-6}
A & B & C & P & Q & R\\  \cline{1-6}
0 & 0 & 0 & 0 & 0 & 0\\ 
0 & 0 & 1 & 0 & 0 & 1\\
0 & 1 & 0 & 0 & 1 & 0\\
0 & 1 & 1 & 0 & 1 & 1\\
1 & 0 & 0 & 1 & 0 & 0\\
1 & 0 & 1 & 1 & 1 & 0\\
1 & 1 & 0 & 1 & 0 & 1\\
1 & 1 & 1 & 1 & 1 & 1\\
\hline
\end{tabular}
\caption{Truth table for the parity preserving Fredkin reversible gate, where  A$\oplus$B$\oplus$C = P$\oplus$Q$\oplus$R}
\label{fig:fredkinGateTruthTable} 
\end{figure}

Figure ~\ref{fig:fredkinGateTruthTable} shows the state table for a 3x3 Fredkin gate, where it can be clearly seen the one-to-one mapping between the inputs and outputs. In addition, Fredkin gate also preserves parity. 

As expected, synthesis of reversible circuits does not follow the traditional methods. First of all, as natural consequence of the one-to-one mapping characteristic of reversible gates, neither feedback loops or fan-outs are allowed. Each gate output is used as an input to the next gate, resulting in a network of cascading gates. In the most brutal fashion, converting an irreversible function into a reversible function is simple; one must simply add extra outputs and inputs to guarantee a one to one mapping. This approach is dis-encouraged as it would unnecessarily increase the size of the design and it would also create many instances where information is not re-used. The extra outputs that are produced to maintain the reversibility of the gate, that are not used downstream are called ~\emph{garbage outputs}. These ~\emph{garbage outputs} are undesirable as they dissipate energy into the circuit. If the maximum number of identical output vectors is ~\emph{p}, then ~\emph{log2 p} garbage outputs (and some inputs) must be added to make the input-output vector mapping unique ~\cite{jhaDate04}. 

To date there has been no definite answer in the synthesis of reversible circuits. Earlier synthesis heuristic methods were plagued by scalability faults and they made extensive use of template matching ~\cite{Miller2003,Iwama2002,Dueck04}. Some of these algorithms would not converge or the extensive use of \emph{garbage outputs} made the designs impractical. In 2006 Gupta et al ~\cite{gupta06} presented a synthesis algorithm using networks of n-bit Toffoli gates and it was expanded in \cite{jha08} to include n-bit Fredkin and Peres gates. The source code of this project has been released to the scientific community since then~\cite{ReLos07}. Other efforts consisted of synthesising using just CNOT gates ~\cite{Yang06} and creating SAT based synthesis with Toffoli gates network synthesis ~\cite{DanielGLSVLSI07}. While recent developments have been promising they are still far from the optimal results computed by slow exhaustive search algorithms ~\cite{Shende02reversiblelogic}.

Despite the fact that at the present time no truly reversible architectures have been successfully manufactured, it is not a far fetched assumption that in the future they will have to rely on individual components prone to failures through faulty components. In reversible architectures, fault tolerance is even more critical than in conventional circuits as errors generate heat, the very problem that the architecture is trying to solve ~\cite{Boykin05}. This work will investigate the possibility of using natural, or artificial, logic invariant relationships in reversible circuits as a method of achieving fault tolerance.

\vspace*{-2mm}
\section{Related Work}
\label{sec:background}

The synthesis rules referenced above impose a very strong restriction on reversible fault tolerance implementations. In non-reversible logic circuits it is not uncommon to use redundant circuit elements which may include simple duplication of the entire design, triple modular redundancy (TMR). This can be trivially implemented in reversible logic with the addition of extra hardware to accommodate input/output fan-out to feed multiple three copies of the circuit, and a reversible comparator ~\cite{Boykin05}. The problem with this approach is the extensive hardware overhead.  Another widely studied approach error consists on the implementation of parity codes. This method has been widely studied in reversible architectures. Since reversible gates have the same number of inputs and outputs, a sufficient requirement for parity preservation of a reversible circuit is that each individual gate be parity preserving, ie. have the same parity for inputs and outputs~\cite{Parhami06}. As an example, Fredkin gate preserves parity (figure ~\ref{fig:fredkinGateTruthTable}) but the Toffoli gate does not. During circuit operation, if we know the parity of the inputs and outputs, we can compare them to see if there is a error caused mismatch. Parhami demonstrated the feasibility of the parity preserving design adding parity protection to Toffoli gates and designing a full adder~\cite{Parhami06}. Others have taken this idea even further by also implementing simple circuits that support error correction with hamming codes ~\cite{Shahana07}. Another error detection method reported in the literature consisted on self checking reversible gate pairs~\cite{Vasudevan05}. While an interesting method the excessive number of garbage outputs made its usefulness very limited. In non-reversible architectures a recent approach to online error detection using naturally occurring invariant relationships has been proposed ~\cite{nepalITC08}. In this paper we will study the feasibility of using invariant relationships in reversible circuits. In addition we will study situations where additional gates are inserted in order to create new invariant relationships.     

\vspace*{-2mm}
\section{Methodology}
\label{sec:methodology}

\begin{figure}
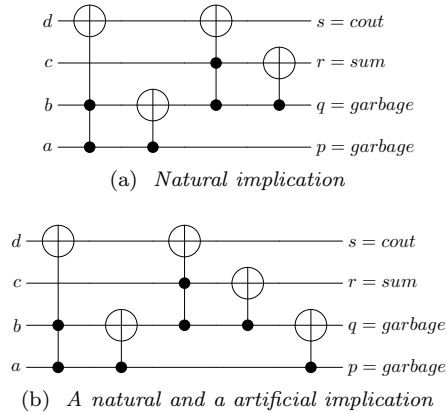

  \centering
  \subfloat[~\emph{Natural implication}]{\includegraphics[scale=0.7]{paper.5}} 
      
  \subfloat[~\emph{A natural and a artificial implication}]{\includegraphics[scale=0.7]{paper.6}}    
    \caption{Invariant relationships in a reversible full adder circuit (rd32) made from Toffoli gates}
  \label{fig:ReversibleAdder}
\end{figure}

An invariant relationship in a circuit happens whenever two sites are expected to hold the same logic value when the circuit is free of errors. Identifying any violation of these invariants could then be used as a means of error detection for the circuit. As an example, consider the CNOT circuit description of a full adder circuit (rd32) Figure~\ref{fig:ReversibleAdder}$a)$. It can be verified that whenever wire $a$ contains $0$ or $1$ at the input, the output logic value will always be $0$ or $1$.  If this relationship does not hold, an error must have occurred in the intervening logic between the two sites or at the second site. These type of implications that naturally occur in the un-modified reversible circuit we call them ~\emph{natural implications}. Due to the synthesis rules outlined above, in reversible logic these invariant relationships can only be implemented at the input/output level. These synthesis rules constraint the number, and quality, of $natural$ invariant relationships in the circuit. One will also notice that with the addition of extra reversible gates one can force the additional creation of invariant relationships. For example, as shown in Figure~\ref{fig:ReversibleAdder}$b)$, when we append a 2x2 Toffoli gate to the unmodified rd32 circuit, with the function $(p,q) \rightarrow (p,p \oplus q)$, we artificially generate another invariant relationship, $(b=0/1) \implies (q=0/1)$. To this type of artificially created logic invariant relationships we call them ~\emph{artificial implications}. It is important to mention that these ~\emph{artificial implications}, can only be placed in locations whose logic output is functionally irrelevant ~\emph{(garbage outputs)}, as they alter the values of the function we are trying to compute. The biggest advantage to these ~\emph{artificial implications} is that, previously useless states now have the purpose of providing circuit fault tolerance.

The procedure to find both ~\emph{natural and artificial implications} is as follows. After choosing a reversible circuit, we run circuit simulation and record the logic state of all output nodes for each input vector applied. In order to avoid potential sampling situations associated with random (or ATPG) vectors simulation, we use all  possible combinations of inputs vectors. For this project, the used circuit simulation tool,~\emph{BRevSim}~\cite{brevsim08}, was custom made as none of the existing publicly available simulators~\cite{WGT+:2008,MDS:2005}, was open-source, supported complex reversible gates or performed fault coverage. Once the simulation is complete, we identify in-variate relationships by comparing node pair values in the simulation runs. For instance, for the $(a=1) \implies (b=1)$ relationship to exist, there should never be an instance where $a=1$ and $b=0$. Since we are exhaustively running all input vectors, we do not need a more formal method to verify that these invariant relationships are valid. With the location of these invariant relationships we have found the ~\emph{natural implications}. 

To create the ~\emph{artificial implications}, first we need to assert which of the circuit wires will contain garbage outputs. Then we need to determine the possibility of appending any reversible gate, in order to establish a new invariant relationship, without altering any non-garbage output logic states. This is accomplished by repeating the same steps used to find ~\emph{natural implications} with random reversible gates being inserted on all permutations of garbage wires. This algorithm is shown on figure~\ref{fig:artificial-imp-algo}. In our work we only used the four common logic gates listed in figure~\ref{fig:sampleRevGates}, but theoretically any reversible gate could be used.

\begin{figure}
  \begin{algorithmic}
    \STATE $test gate\gets $ any reversible gate
    \FOR{each garbage wire permutation} 
    \STATE - place $test gate$ on wire permutation
    \STATE - run circuit simulation
    \STATE - check for ~\emph{artificial implications}
    \ENDFOR
  \end{algorithmic}
  \caption{Exhaustive search algortihm to find ~\emph{artificial implications}}
  \label{fig:artificial-imp-algo}
\end{figure}

Also, in our model we do not allow the attached gate to be in any non-garbage producing wire, even if the output for that particular line is not modified. In the Toffoli Gate outlined on Figure~\ref{fig:sampleRevGates}, we can see that in the first wire, the output and input value of the gate remain unchanged, therefore we could technically allow that particular wire to be on a non-garbage output. However we chose not to do so. The reason being that any inserted gate has the potential to fail, and a failure in a garbage wire will be of no consequence, while a failure elsewhere could be catastrophic.  

Once we've determined the ~\emph{natural and artificial implications}, we performed fault simulation in order to assert the impact of each invariant relationships. We would like to know, for example, how many reversible faulty components could theoretically be detected by our ~\emph{implications}. Similar to the work in ~\cite{nepalITC08}, we would like an invariant relationship to be able to detect as many faults in the circuit as possible.  The reversible fault model used in this work is the reversible implementation stuck-at-0/stuck-at-1 outlined in existing work~\cite{PolianATS05}. The model suggests the insertion of a fault on every wire before each gate, even if a particular wire does not impact the outcome of the reversible gate. According to this model, the total number of faults for a given circuit is $number of gates * number of wires * 2$. With this model, the circuit rd32 listed on figure~\ref{fig:ReversibleAdder} has 32 faults, half of them being stuck-at-zero. The degree of accurateness of this fault model is beyond the scope of this work. Considering that no reversible circuit has been produced, there can be no consensus whether this fault model is less accurate than other proposed reversible fault models ~\cite{ZhongEvolutionary2006,Patel04faulttesting}. We opted for this fault model, because we had to choose one, and this particular one would generate an excess of faults guaranteeing good stress test to our reversible circuit simulation platform.

It is also beyond the scope of this project the implementation, or conception, of the hardware required to assert the validity of each implication during circuit run-time.

\vspace*{-2mm}
\section{Results}
\label{sec:results}

\begin{table}
  \begin{center}
    \begin{tabular}{ | l | c | c | c | c |}
      \hline
      benchmark  & \# gates  &  \# wires      & \# garbage & source \\ \hline
      rd32       & 4         &  4             & 2           & ~\cite{MDS:2005}  \\ 
      rd53-130   & 30        &  7             & 4           & ~\cite{WGT+:2008} \\ 
      rd84-143   & 21        &  15            & 11          & ~\cite{WGT+:2008} \\ 
      sym6-145   & 36        &  7             & 6           & ~\cite{WGT+:2008} \\
      4gt4-v0-73 & 17        &  5             & 4           & ~\cite{WGT+:2008} \\ 
      alu-v4-6   & 7         &  5             & 4           & ~\cite{WGT+:2008} \\ 
      9symd2     & 28        &  12            & 11          & ~\cite{MDS:2005}  \\
      ckt1-149   & 11553     &  9             & 0           & ~\cite{WGT+:2008} \\
      ham7-25-49 & 25        &  7             & 6           & ~\cite{MDS:2005}  \\
      hwb6-56    & 126       &  6             & 0           & ~\cite{WGT+:2008} \\
      \hline
    \end{tabular}
  \end{center}  
  \caption{Characteristics of the reversible of benchmarks used on our simulations}
  \label{tab:BenchmarksTable} 
\end{table}

On table~\ref{tab:BenchmarksTable} we list the selected reversible benchmarks to be used on our analysis. Unfortunately most of the available benchmark circuits are either very small or they have a strong predilection to being synthesized with Toffoli gates. In the same table, the \emph{number of wires} correspond to the number of input/output, and the ~\emph{number of garbage} correspond to the number of the number of those wires whose logic value is irrelevant on the circuit. 

Table~\ref{tab:ImpTable} contains the number of discovered ~\emph{natural and artificial implications}. A quick glimpse at the table shows us that there are very few natural implications for larger and more complex circuits with no garbage outputs, such as ~\emph{hwb6-56} or ~\emph{ckt1-149}. In fact, except for ~\emph{ham7-25-49} there is a visible relationship between the number of garbage outputs and the detected implications. We then looked individually at each ~\emph{natural implication} detected, and we noticed that every single one of them existed due to a wire whose logic value was never altered, the same behaviour outlined on figure~\ref{fig:ReversibleAdder}a). This result was disheartening as during the conception of this work we were expecting to find more complex implications. The detection of ~\emph{artificial implications} was also extremely disappointing. Even with exhaustive search, the only circuit, besides the already studied ~\emph{rd32}, that gave us some new ~\emph{artificial implications} was ~\emph{9symd2}. Unfortunately  every single one of ~\emph{artificial implications} was generated due to a trivial addition of a 2x2 Toffoli gate, the same type of ~\emph{artificial implication} described on figure~\ref{fig:ReversibleAdder}b). The other circuits did not yield any new implications either because there were no more new implications to detect, or not enough garbage wires. 

Despite these poor results, we were still curious about the fault coverage impact of each individual implication on the overall circuit. To the proposed metric we call ~\emph{implication impact} and it is the ratio between the number of faults detected at the output that caused an implication to be violated and the number of faults that were detected at the output. The algorithm for metric is described in Figure ~\ref{fig:impImpact}, and  obviously the higher the implication impact the better. A 55\% implication implicat would tell us that a particular implication can be expected to detect faults that propagated to the output, an estimated 55\% of the times.

\begin{table}[ht!]
  \begin{center}
    \begin{tabular}{|l|c|c|c|c|} \hline
      \multicolumn{1}{|c}{} & \multicolumn{2}{|c|}{natural implications} & \multicolumn{2}{|c|}{artificial implications}  \\ \cline{2-5}
      benchmark  & number    & avg. impact    & number      & avg. impact \\ \hline
      rd32       & 1         &  12.5          & 1           & 18.75\\
      rd53-130   & 3         &  7.14          & 0           & 0\\
      rd84-143   & 1         &  0             & 0           & 0\\
      sym6-145   & 5         &  5.12          & 0           & 0\\
      4gt4-v0-73 & 0         &  0             & 0           & 0\\
      alu-v4-6   & 1         &  10            & 0           & 0\\
      9symd2     & 2         &  8.2           & 7           & 22.5\\
      ckt1-149   & 0         &  0             & 0           & 0\\
      ham7-25-49 & 0         &  0             & 0           & 0\\
      hwb6-56    & 0         &  0             & 0           & 0\\

      \hline
    \end{tabular}
  \end{center}
 \caption{Number of different type of implications and their impact}
  \label{tab:ImpTable}
\end{table}

\begin{figure}
  \begin{algorithmic}
    \STATE $testImplication \gets $ any valid implication
    \STATE $errorDetected \gets $ 0
    \STATE $errorMissed \gets $ 0

    \FOR{each input vector} 
    \FOR{each fault} 
    \IF{$TestImplication$ was violated $\bigwedge$ error was propagated to output}    
    \STATE $errorDetected \gets $ $errordetected ++ $
    \ENDIF
    \IF{$TestImplication$ was NOT violated $\bigwedge$ error was propagated to output}    
    \STATE $errorMissed \gets $ $errorMissed ++ $
    \ENDIF
    \ENDFOR
    \ENDFOR
    \STATE $implicationImpact \gets \frac{errorDetected}{errorMissed+errorDetected} * 100 $
  \end{algorithmic}
\caption{Algorithm that calculates the implication impact}
  \label{fig:impImpact}
\end{figure}

As it can be seen on table~\ref{tab:ImpTable}, not only we have very little implications, but these also have very little overall impact. The results were so poor that we did not see the need to run a circuit simulation with every single detected implication. One point worth noting is that the average impact of the detected ~\emph{artificial implications} is greater than the average impact of the ~\emph{natural implications}, that is mainly due to the fact that we added an extra gate at the output which allowed us to detect more faults. It is worth noting that $rd84$ only has one detected implication with a 0\% impact. This indicated that all the faults that could be detected by the implication circuitry are the faults that are \textbf{not} propagated to the output. 

Finally, in terms of computational performance, our experiments have shown the small size of the benchmark reversible circuits, performing an exhaustive input vector simulation took a negligible time. For a sample circuit with 10 input lines and 50 gates, circuit simulation in ~\emph{BRevSim} takes less than 1 second on a Dual Core 2GHz 2007 MacBook computer.

\section{Conclusions}
\label{sec:conclusions}
The purpose of this paper was to assert if invariant relationship were also viable for online error detection in reversible circuits. our results show that unlike non reversible circuits, the implication concept is ineffective. We observed that when implications are naturally present in the reversible circuit they tend to be very simple in nature, occurring due to wires with unchanged logic states. These uninteresting implications obviously cover few faults. When we recursively searched for ~\emph{artificial implications}, we found that they rarely existed, even with exhaustive search mechanisms. When these exist, these were trivial in nature. In conclusion, our simulations on a small subset of reversible benchmarks, strongly indicate that implications are \textbf{not} suited for reversible circuits.

\section{Future Work}
\label{sec:futurework}
One of the positive outcomes of this paper is that a reversible circuit simulation capable of fault analysis was released as an open source project. With it several extensions to this paper can be researched, one example being the investigation of reversible logic synthesis algorithms so that unavoidable garbage outputs, should contain logic implications. Another potential research area direction consist on actually designing the hardware that detects the violations of the invariant relationships.

\bibliography{references}
\bibliographystyle{abbrv}

\end{document}